\documentclass[prb,reprint,showpacs]{revtex4-1}

\usepackage{hyperref}   
\usepackage[utf8]{inputenc}
\usepackage{multirow}
\usepackage{tabulary}
\usepackage{siunitx}
\usepackage{threeparttable}

\usepackage[T1]{fontenc}
\usepackage[english]{babel}
\usepackage{booktabs}
\usepackage{tikz}
\usepackage{titlesec}
\usepackage{threeparttable}				
\usepackage{appendix}
\usepackage{MnSymbol}
	\usepackage{hyperref}
	\hypersetup{
		colorlinks,
		citecolor=blue,
		linkcolor=blue,
		urlcolor=blue}

\newcolumntype{K}[1]{>{\centering\arraybackslash}p{#1} }

\begin{document}

\title{Investigation of potassium-intercalated bulk MoS$_2$ using transmission electron energy-loss spectroscopy}
\author{Carsten Habenicht}
\email{c.habenicht@ifw-dresden.de}
\author{Axel Lubk}
\author{Roman Schuster}
\author{Martin Knupfer}
\author{Bernd Büchner}
\affiliation{IFW Dresden, Institute for Solid State Research, Helmholtzstrasse 20, 01069 Dresden, Germany}
\date{\today}

\begin{abstract}
We have investigated the effect of potassium (K) intercalation on $2H$-MoS$_2$ using transmission electron energy-loss spectroscopy. 
For K concentrations up to approximately 0.4, the crystals appear to be inhomogeneous with a mix of structural phases and irregular potassium distribution. Above this intercalation level, MoS$_2$ exhibits a $2a \times 2a$ superstructure in the $ab$ plane and unit cell parameters of $a=\SI{3.20}{\angstrom}$ and $c=\SI{8.23}{\angstrom}$ indicating a conversion from the $2H$ to the $1T'$ or $1T''$ polytypes. The diffraction patterns also show a $\sqrt{3}a \times \sqrt{3}a$ and a much weaker $2\sqrt{3}a \times 2\sqrt{3}a$ superstructure that is very likely associated with the ordering of the potassium ions. A semiconductor-to-metal transition occurs signified by the disappearance of the excitonic features from the electron energy-loss spectra and the emergence of a charge carrier plasmon with an unscreened plasmon frequency of $\SI{2.78}{\eV}$. The plasmon has a positive, quadratic dispersion and  appears to be superimposed with an excitation arising from interband transitions.
The behavior of the plasmon peak energy positions as a function of potassium concentration shows that potassium stoichiometries of less than $\sim 0.3$ are thermodynamically unstable while higher stoichiometries up to $\sim 0.5$ are thermodynamically stable. Potassium concentrations greater than $\sim 0.5$ lead to the decomposition of MoS$_2$ and the formation of K$_2$S.
The real part of the dielectric function and the optical conductivity of K$_{0.41}$MoS$_2$ were derived from the loss spectra via Kramers-Kronig analysis.

\end{abstract}

%\pacs{79.20.UV, 71.35.-y,73.21.Ac}

\maketitle

\section{INTRODUCTION}
MoS$_2$, a semiconducting transition metal dichalcogenide (TMDC), forms quasi-two-dimensional, layered crystals. The weak interlayer bonding forces permit the introduction of intercalants between the layers. Those intercalants can affect the properties of the host material in ways that make it useful for obtaining deeper insight into fundamental physical processes as well as technical applications.
Because of their low electron affinity, alkali metals are well suited as electron donors when inserted into TMDC. So far, the majority of the research efforts in that respect were directed towards lithium intercalated MoS$_2$. However, potassium-doped molybdenum disulfide has also received a significant amount of attention. Among other things, the compound is of interest because of its potential to form structurally different polytypes accompanied by a significant change of its electronic properties depending on the potassium (K) loading. That makes it a good platform for engineering specific electronic and structural characteristics. For example, its K-doped $2H$ polytype was found to be superconducting with critical temperatures of $T_c=4.5-\SI{6.9}{\K}$\cite{Somoano_PhysicalReviewLetters_1971_27_7_402, Somoano_TheJournalofChemicalPhysics_1973_58_2_697, Woollam_PhysicalReviewB_1976_13_9_3843, Zhang_Nl2015_629} while the $1T$ phases have lower transition temperatures of $2.8-\SI{4.6}{\K}$.\cite{Zhang_Nl2015_629, Guo_JoMCC2017_10855, Fang_ACIE2018_1232} 
From a practical point of view, $2H$ has some catalytic properties that make it useful for hydrogen evolution reactions.\cite{Li_JACS2011_7296, Wang_PotNAoS2013_19701, Tsai_Nl2014_1381} However, the $1T$ and $1T'$ phases are considerable more efficient\cite{Maitra_ACIE2013_13057, Lukowski_JACS2013_10274, Wang_PotNAoS2013_19701, Voiry_NL2013_6222, Chou_Nc2015_8311, Putungan_PCCP2015_21702, Bai_NR2015_175, Laursen_EES2012_5577, Fan_JMCA2014_20545, Gao_JPCC2015_13124} and can be evoked by K-doping.

Moreover, the intercalation can also induce a semiconductor-to-metal transition due to the infusion with additional electrons and the changes in the electronic structure which makes it a promising anode material for potassium-based batteries.\cite{Ren_NR2017_1313} That transition made it an interesting material for an investigation using transmission electron energy-loss spectroscopy (EELS). This method permits not only the momentum dependent measurement of single particle excitations but also collective density oscillations of charge carriers, referred to as plasmons,\cite{Nozieres_PR1959_1254, Raether_2006_} which are typically associated with metals. It has been used successfully for other materials\cite{Nuecker_PRB1991_7155, Wezel_PRL2011_176404} and in particular for  intercalated TMDCs.\cite{Koenig_EL2012_27002, Koenig_PRB2013_195119, Mueller_PRB2016_35110, Ahmad_JoPCM2017_165502} We employed EELS to pristine and K-intercalated MoS$_2$ to gain insight into the change of the crystal structure as well as the semiconductor-to-metal transition leading to the development of a charge carrier plasmon, the collective excitation of all conduction electrons.
\begin{figure*}
	\includegraphics [width=\textwidth]{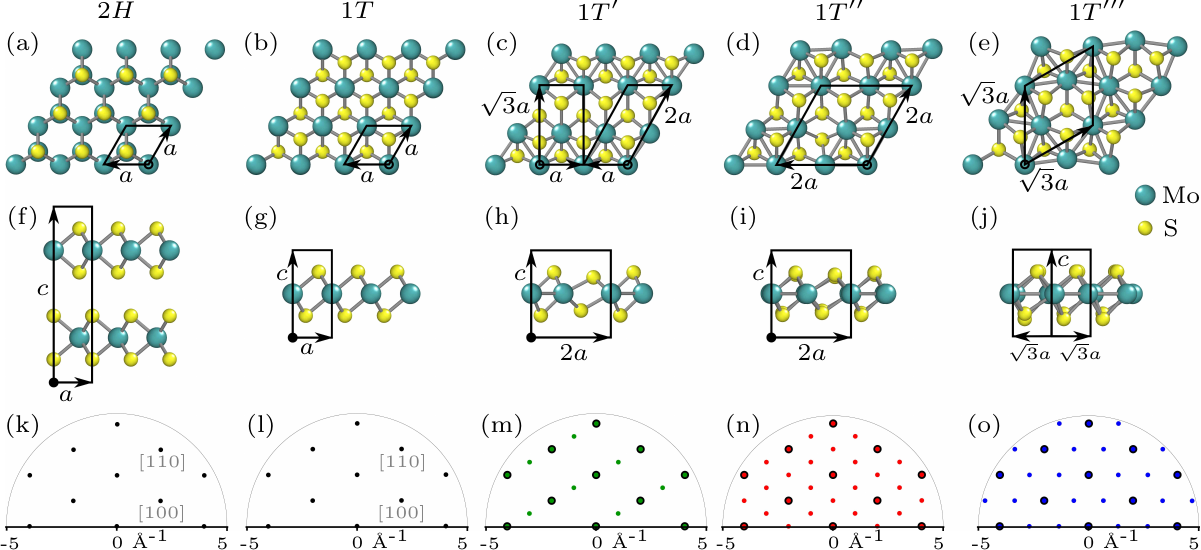}
	\caption{Crystal structures and unit cells for the in-plane [(a)-(e)] and out-of-plane directions [(f) - (j)] for the polytypes of MoS$_2$ indicated in the first row. (k)-(o) show the simulated in-plane electron diffraction patterns for the respective polytypes. For a better comparison of the diffraction patterns, the diffraction peaks for the distorted polytypes whose locations agree to the ones from the undistorted material are depicted as larger black circles in (m)-(o).}
	\label{fig:CrystalStructures}
\end{figure*}
\section{MoS$_2$ polymorphism}
MoS$_2$ forms crystals made up of parallel stacked layers. Each of them is composed of an atomically thin sheet of molybdenum atoms sandwiched between two sheets of sulfur atoms. The atoms within a layer are held together by covalent bonds while the layers among each other are joined by comparatively weak Van der Waals forces.\cite{Heda_JournalofPhysicsandChemistryofSolids_2010_71_3_187, Bronsema_ZfuauaC1986_15}
The material assumes a number of polytypes. Under typical ambient conditions, the most stable\cite{Chou_Nc2015_8311} and, therefore, most commonly found polytype is $2H$-MoS$_2$ (space group: $P6_3/mmc$ [194]). In that case, six chalcogen atoms are arranged in a trigonal prismatic coordination around a metal atom.\cite{Dickinson_JournaloftheAmericanChemicalSociety_1923_45_6_1466, Wilson_AdvancesinPhysics_1969_18_73_193, Bronsema_ZfuauaC1986_15, Friend_AiP1987_1} Each unit cell comprises two layers that are shifted in respect to each other such that the molybdenum atoms in one layer are collinear in the out-of-plane direction with the sulfur atoms in the adjacent layers [see Figs. \ref{fig:CrystalStructures}(a) and \ref{fig:CrystalStructures}(f)]. The material is semiconducting and has an indirect band gap of $1.2-\SI{1.3}{\eV}$ in its bulk form.\cite{Jiang_TheJournalofPhysicalChemistryC_2012_116_14_7664, Baglio_JournaloftheElectrochemicalSociety_1982_129_7_1461, Cheiwchanchamnangij__2014___, Mattheiss_PhysicalReviewB_1973_8_8_3719, Komsa_PhysicalReviewB_2012_86_24_241201}

Besides the $3R$\cite{JELLINEK_N1960_376} and $1H$\cite{Calandra_PRB2013_245428, Fan_JMCA2014_20545} phases, which are not relevant for this work, four additional phases have been described where the arrangement of the sulfur atoms around the molybdenum atom is octahedral instead of trigonal prismatic.\cite{Py_CJoP1983_76} Their unit cells span only one molecular layer. One member of this family is the undistorted $1T$ phase\cite{Fang_ACIE2018_1232} [space group: $P\bar{3}m1$ [164], Figs. \ref{fig:CrystalStructures}(b) and \ref{fig:CrystalStructures}(g)]. 
First principle calculations for monolayers\cite{Shirodkar_Prl2014_157601,Fan_JMCA2014_20545, Kan_TJoPCC2014_1515, Gao_JPCC2015_13124, He_NRL2016_330, Pal_PRB2017_195426} and experiments on 1-3 layer samples found it to be metallic.\cite{Kappera_Nm2014_1128}
There are three more distorted structures. In $1T'$, the metal atoms form parallel running zig-zag chains leading to in-plane lattice parameters of $2a \times a$ which can also be described by an orthorhombic $\sqrt{3}a \times a$ cell [see Figs. \ref{fig:CrystalStructures}(c) and \ref{fig:CrystalStructures}(h)].\cite{Heising_JotACS1999_638, Gordon_PRB2002_125407, Eda_AN2012_7311, Calandra_PRB2013_245428, Fan_JMCA2014_20545, Guo_JoMCC2017_10855} Here $a$ refers to the lattice parameter of undistorted $1T$-MoS$_2$ which is approximately that of the $2H$ polytype.\cite{Fan_JMCA2014_20545} Some calculations predict a band gap between 0.006 and $\SI{0.26}{\eV}$ for monolayers\cite{Fan_JMCA2014_20545, Kan_TJoPCC2014_1515, Qian_S2014_1344, Chou_Nc2015_8311, Gao_JPCC2015_13124, Liu_JoPCM2017_95702, Pal_PRB2017_195426} while others forecast metallic behavior\cite{Singh_2M2015_35013}.
In the $1T''$-form, the molybdenum atoms tetramerize in diamond-like clusters in the $ab$ plane and the unit cell parameters become $2a \times 2a$ [see Figs. \ref{fig:CrystalStructures}(d) and \ref{fig:CrystalStructures}(i)].\cite{Yang_PRB1991_12053, Calandra_PRB2013_245428} First-principal calculations indicate that it is metallic.\cite{Pal_PRB2017_195426} It should be pointed out that Chou \textit{et al.} use the notation $1T''$ to refer to a different structure with trimerized Mo pockets. 
\cite{Chou_Nc2015_8311}
Bulk $1T'''$ exhibits trimerized metal atoms each of them being the corner of two adjacent trimers.\cite{Fang_JotACS2019_790} Its space group is $P3_1m$ forming a $\sqrt{3}a \times \sqrt{3}a$ superstructure [see Figs. \ref{fig:CrystalStructures}(e) and \ref{fig:CrystalStructures}(j)].\cite{Wypych_JCSCC1992_1386, Wypych_Com1998_723,Wypych_JoSSC1999_430, Shirodkar_Prl2014_157601, Fang_JotACS2019_790} Resistivity and absorption measurements found this polytype to be semiconducting with a band gap of $\SI{0.65}{\eV}$\cite{Fang_JotACS2019_790}, which was confirmed by \textit{ab initio} calculations\cite{Shirodkar_Prl2014_157601}, while an earlier experiment\cite{Wypych_JCSCC1992_1386} determined it to be metallic without providing supporting data. Monolayers of $1T'''$ have a band gap of about $\SI{0.7}{\eV}$.\cite{Shirodkar_Prl2014_157601, Bruyer_PRB2016_195402} 

In $2H$-MoS$_2$, the starting material of our investigation, the states closest to the Fermi level are largely represented by Mo$^{4+}$ $4d$ states. In terms of ligand field theory, the two unoccupied conduction bands are formed by $4d_{xy,x^2-y^2}$ and $4d_{xz,yz}$ orbitals, respectively. The $4d_z^2$ orbital makes up the valance band and is fully occupied with two electrons resulting in a structurally stable semiconductor.\cite{Enyashin_TJoPCC2011_24586, Enyashin_CaTC2012_13}
This system can be destabilized via infusion of additional electrons by methods such as intercalation with electron donors, plasmonic hot electrons doping\cite{Kang_AM2014_6467}, substitutional doping (e.g rhenium)\cite{Enyashin_TJoPCC2011_24586, Lin_Nn2014_391} and electron beam exposure\cite{Lin_Nn2014_391}. Intercalation with alkali metals (e.g. lithium, sodium, potassium) as electron donors has been performed by electrochemical means\cite{Dahn_CJoP1982_307, Mulhern_Cjop1989_1049, Imanishi_SSI1992_333, Cheng_An2014_11447, Huang_AMI2017_1700171,Ren_NR2017_1313}, wet-chemical methods\cite{Benavente_CCR2002_87, Ruedorff_Chimia_1965_19__489} or thermal evaporation of the intercalant\cite{Somoano_TheJournalofChemicalPhysics_1973_58_2_697}.   
The additional electrons will occupy a previously empty $d$-orbital above the Fermi level making the material structurally unstable.\cite{Enyashin_TJoPCC2011_24586, Enyashin_CaTC2012_13} This causes the transition from the $2H$ to the $1T$ polytype in which a rearrangement of the orbital occurs. In the new phase, the Fermi level passes between the $4d_{xy,xz,yz}$ ($t_{2g}$) orbitals and the empty $4d_{z^2, x^2-y^2}$ ($e_g$) states.\cite{Kertesz_JACS1984_3453, Enyashin_TJoPCC2011_24586, Enyashin_CaTC2012_13,Chhowalla_Nc2013_263, Voiry_CSR2015_2702, Gao_JPCC2015_13124} Scanning transmission electron microscopy allowed the observation of this conversion process which takes place by sliding of the sulfur plans\cite{Lin_Nn2014_391} along a moving front\cite{Gao_AN2015_11296}. Ligand field theory predicts that a donated electron moves into the threefold degenerate $t_{2g}$ states of the reduced Mo$^{3+}$ ions leaving each of the associated three orbitals half-filled. Those orbitals point towards three adjacent Mo atoms forming metal-metal bonds that lead to the formation of diamond-like patterns with a $2a \times 2a$ superstructure [Fig. \ref{fig:CrystalStructures}(n)].\cite{Rocquefelte_PRB2000_2397, Petkov_PhysicalReviewB_2002_65_9_92105, Enyashin_CaTC2012_13}
Li, Na and K-intercalated bulk and few-layer MoS$_2$ actually displayed this kind of superstructure for sufficiently high alkali metal concentrations in experiments.\cite{Chrissafis_MSaEB1989_145, Dungey_CoM1998_2152, Wypych_JoSSC1999_430, Cheng_An2014_11447, Huang_AMI2017_1700171, Huang_SCC2018_222} 

On a microscopic level, however, investigations found the coexistence of domains with $2a \times a$ as well as $2a \times 2a$ superstructures which on a macroscopic level appeared to be only $2a \times 2a$ based on their diffraction patterns.\cite{Wang_JotACS2014_6693, Gao_AN2015_11296} Studies on restacked MoS$_2$ attributed this effect to the twinning of differently oriented crystal domains. The combination of three $2a \times a$ superstructures rotated by $60^{\circ}$ (or $120^{\circ}$) in respect to each other generates a diffraction pattern that is identical to that of a $2a \times 2a$ superlattice\cite{Amelinckx_JPC1976_4, Heising_JotACS1999_638} [see Figs. \ref{fig:CrystalStructures}(m) and \ref{fig:CrystalStructures}(n) for simulated diffraction patterns generated by those two superstructures]. This tripling makes it difficult to distinguish the two lattice structures, and therefore the $1T'$ and $1T''$ phases, based on the diffraction measurements obtained over macroscopic sample volumes. The $2a \times a$ arrangement was also identified in single layer MoS$_2$.\cite{Sandoval_PRB1991_3955, Qin_PRB1991_3490, Qin_U1992_630} Moreover, the observed superstructures appear to depend on the alkali metal concentration.\cite{Mulhern_Cjop1989_1049} Huang \textit{et al.} also found a $2a \times \sqrt{3}$ structure for short sodiation times.\cite{Huang_SCC2018_222}

\section{EXPERIMENT}\label{sec:Experiment}
Using adhesive tape, thin films of approximately $\SI{100}{\nano\meter}$-thickness were prepared by exfoliating bulk molybdenum disulfide along the Van der Waals gaps. Those natural single crystals had been purchased from Manchester Nanomaterials. Under an optical microscope, the transparency of the cleaved films was compared to that of samples cut with a calibrated microtome to estimate film thickness. 
$\SI{100}{\nano\meter}$-samples were selected because they exhibit acceptable EELS count rates, which become too low for thinner films, while reducing the negative effects of multiple scattering, which increase with the number of layers. The samples were transferred to platinum transmission electron microscopy grids and placed into the 172-keV transmission electron energy-loss spectrometer. The instrument is described in more detail in Refs. \onlinecite{Fink_AEEP_1989_75__121, Roth_JournalofElectronSpectroscopyandRelatedPhenomena_2014_195__85}. EELS is a bulk sensitive scattering technique whose spectra are proportional to the loss function $L(\vec{q},\omega)=\text{Im}[-1/\epsilon (\vec{q},\omega)]$ where $\epsilon (\vec{q},\omega)$ is the dielectric function depending on momentum $\vec{q}$ and energy $\omega$.\cite{Sturm_ZNA1993_233} All experiments were performed at a temperature of $\SI{300}{\K}$ under ultrahigh vacuum.

Because the spectrometer works in transmission mode and the electron beam's spot size is on the order of $\SI{1}{\square\mm}$, all measurements represent an average over the exposed sample volume. Consequently, if a sample is non-uniform, the resulting spectra and diffraction patterns are a superposition of the effects from the various domains. 

The films were intercalated by placing them above SAES alkali metal dispensers from which potassium was thermally evaporated in ultrahigh vacuum (base pressure below $\SI[parse-numbers=false]{10^{-10}}{\milli\bar}$). Different intercalation levels were achieved by exposing the samples to the potassium stream for time intervals ranging from $\SI{45}{\second}$ to $\SI{15}{\minute}$ between the EELS measurements. After each doping process, the films were annealed for $\SI{\sim1}{\hour}$ at $80 -\SI{120}{\celsius}$. The practicability of this approach has been established in a number of previous investigations.\cite{Koenig_EL2012_27002, Koenig_PRB2013_195119, Mueller_PRB2016_35110, Ahmad_JoPCM2017_165502}

For the pristine and intercalated samples, we measured the energy-loss spectra in the $\Gamma$M and $\Gamma$K directions of the Brillouin zone (1) in the energy range of $0.2-\SI{70}{\eV}$ for a momentum transfer of $|\vec{q}|=\SI{0.1}{\per\angstrom}$ and (2) the energy region of $0.2-\SI{4}{\eV}$ for various $|\vec{q}|$ values between 0.075 and $\SI{0.5}{\per\angstrom}$. Moreover, electron diffraction pattern for the same directions were acquired for $\SI{0.2}{\per\angstrom}\!<|\vec{q}|<\SI{6.0}{\per\angstrom}$ at $E=\SI{0}{\eV}$ (elastic scattering). For selected intercalation levels, 180 of such diffraction patterns were measured in $\SI{1}{\degree}$ steps from one $\Gamma$K direction to the one pointing in the opposite direction. The 180 data set were combined to form diffraction maps of half of the in-plane Brillouin zone. The momentum and energy resolutions of the employed instrument setup were $\Delta |\vec{q}|=\SI{0.04}{\per\angstrom}$ and $\Delta E=\SI{82}{\milli\eV}$, respectively. Moreover, energy-loss spectra for the core levels of sulfur and potassium were obtained ranging from 125 to $\SI{320}{\eV}$ with $|\vec{q}|=\SI{0}{\per\angstrom}$, $\Delta |\vec{q}|=\SI{0.07}{\per\angstrom}$ and $\Delta E=\SI{369}{\milli\eV}$.

The momentum transfer of elastically scattered electrons can be regarded as lying in a plane perpendicular to the electron beam. The film surface, which corresponds to the $ab$ plane in the case of MoS$_2$, is typically positioned parallel to the momentum transfer plane but can be rotated in a range of up to $\SI{45}{\degree}$ away from it. This angle is subsequently referred to as polar angle which is $\SI{90}{\degree}$ offset from the angle of incidence. 
This setup allows performing diffraction measurements via elastic electron scattering in the $ab$ plane (polar angle $=\SI{0}{\degree}$) but not directly in the $c$ direction. However, information about the out-of-plane-direction can be obtained by changing the polar angle until the reciprocal lattice points of the adjacent crystal lattice layer are aligned with the plane of the momentum transfer allowing the observation of the diffraction peak associated with the neighboring plane. The momentum positions of two diffraction features that are equivalent in the $ab$ plane but not the $c$ direction (e.g. [110] and [111]) can than be related via the Pythagorean theorem to calculate the separation of the layers in momentum space which then can be converted to real space values. This process can be repeated for successive planes (e.g. [110], [111], [112], ...) until the maximum polar angle of $\SI{45}{\degree}$ is reached. That approach allowed us to determine the $a$- as well as the $c$-parameters of the unit cell.

We measured low-momentum energy-loss spectra before and after the long-time dispersion measurements and found no noticeable changes suggesting that the samples did not suffer beam damage. We did not observe a beam induced phase change from the $2H$ to a $1T$ phase, which was a concern based on research in this respect\cite{Lin_Nn2014_391}, nor a reversion of the phase transition due to aging or heating\cite{Yang_PRB1991_12053, Sandoval_PRB1991_3955, Wypych_JCSCC1992_1386, Eda_Nanoletters_2011_11_12_5111, Guo_JoMCC2017_10855}. Because of the ultra high vacuum conditions, it is unlikely that the presence of H$_2$O affected the experimental results. 

To check for the formation of K$_2$S due to the intercalation, comparison spectra of potassium sulfide were taken. For that purpose, K$_2$S was purchased from Strem Chemicals, Inc. (CAS number: 1312-73-8, purity$\geq$ 95\%). The salt was dissolved in ethanol in a nitrogen atmosphere inside a polyethylene glove bag. The solution was dispensed onto platinum transmission electron microscopy grids one drop at a time. After each drop, the solvent was allowed to evaporate so a salt film could form on the grid. The K$_{2}$S-covered grids were transferred to the spectrometer in a nitrogen-filled vacuum suitcase to prevent contact with the ambient atmosphere. 

Kramers-Kronig analysis was applied to some of the loss spectra of intercalated MoS$_2$. The obtained optical conductivity function and the real part of the dielectric function were fitted based on the Drude-Lorentz model. It describes the complex dielectric function $\epsilon(\omega)$ as a function of the frequency $\omega$ of a series of oscillators modeling the excitation of the involved particles in a material:\cite{Rakic_Appliedoptics_1998_37_22_5271, Egerton_Electronenergy-lossspectroscopyintheelectronmicroscope_2011}
\begin{subequations}
\begin{align}
\epsilon(\omega)&= \epsilon_{r}(\omega)+i\epsilon_{im}(\omega), \label{equ:E_Total}\\
\epsilon_{r}(\omega)&=\underbrace{\epsilon_{\infty}-\frac{\omega_{p}^2}{\omega^2+\gamma^2}}_\text{Drude term}+\underbrace{
\sum_{j=1}\omega_{pj}^2\frac{\omega_j^2-\omega^2}{\left(\omega_j^2-\omega^2 \right)^2+\gamma_j^2 \omega^2}}_\text{Lorentz terms},\label{equ:E_Real}\\
\epsilon_{im}(\omega)&=\underbrace{\frac{\omega_{p}^2\gamma}{\omega(\omega^2+\gamma^2)}}_\text{Drude term}+\underbrace{\sum_{j=1}\omega_ {pj}^2\frac{\gamma_j \omega}{\left(\omega_j^2-\omega^2 \right)^2+\gamma_j^2 \omega^2}}_\text{Lorentz terms}. \label{equ:E_Imag}
\end{align}
\end{subequations}
$j$ is the index number of each oscillator included in the computations and $\omega_j$ the resonant frequency of the $j$-th oscillator. $\gamma$ and $\gamma_j$ are the damping factors (frequency width) of the Drude and $j$-th Lorentz oscillators, respectively. $\epsilon_{\infty}$ symbolizes the background dielectric constant to account for the effect of oscillators not explicitly included in the sum.      
$\omega_{p}$ ($\omega_{pj}$), the plasmon frequency expressing the oscillator strength, can be further broken down:
\begin{subequations}
\begin{flalign}
	\text{for the Drude term:\;}\omega_{p}=\sqrt{\frac{q_0^2 n}{m_{e}\epsilon_0}}\;\text{and} \label{equ:PlasmonFrequDrude}\\
	\text{for the Lorentz terms:\;}\omega_{pj}=\sqrt{\frac{q_0^2 n_j}{m_{ej}\epsilon_0}}\label{equ:PlasmonFrequLorentz}
\end{flalign}
\end{subequations}
with $q_0$ being the elementary electron charge, $\epsilon_0$ the permittivity of free space, $n$ ($n_j$) the number of electrons per unit volume and $m_e$ ($m_{ej}$) the effective electron mass associated with the Drude oscillator (the $j$-th Lorentz oscillator).

The discussion of the method used to calculate the intercalation levels is deferred to Sec. \ref{sec:IntercalationLevel} because it is based on a significant portion of the experimental findings. Therefore, it is more practical to present the experimental results first.

\section{Results and Discussion}\label{sec:R&D}
\subsection{Structural phase conversion}\label{sec:CrystalParameters}
\begin{figure*}
	\includegraphics [width=\textwidth]{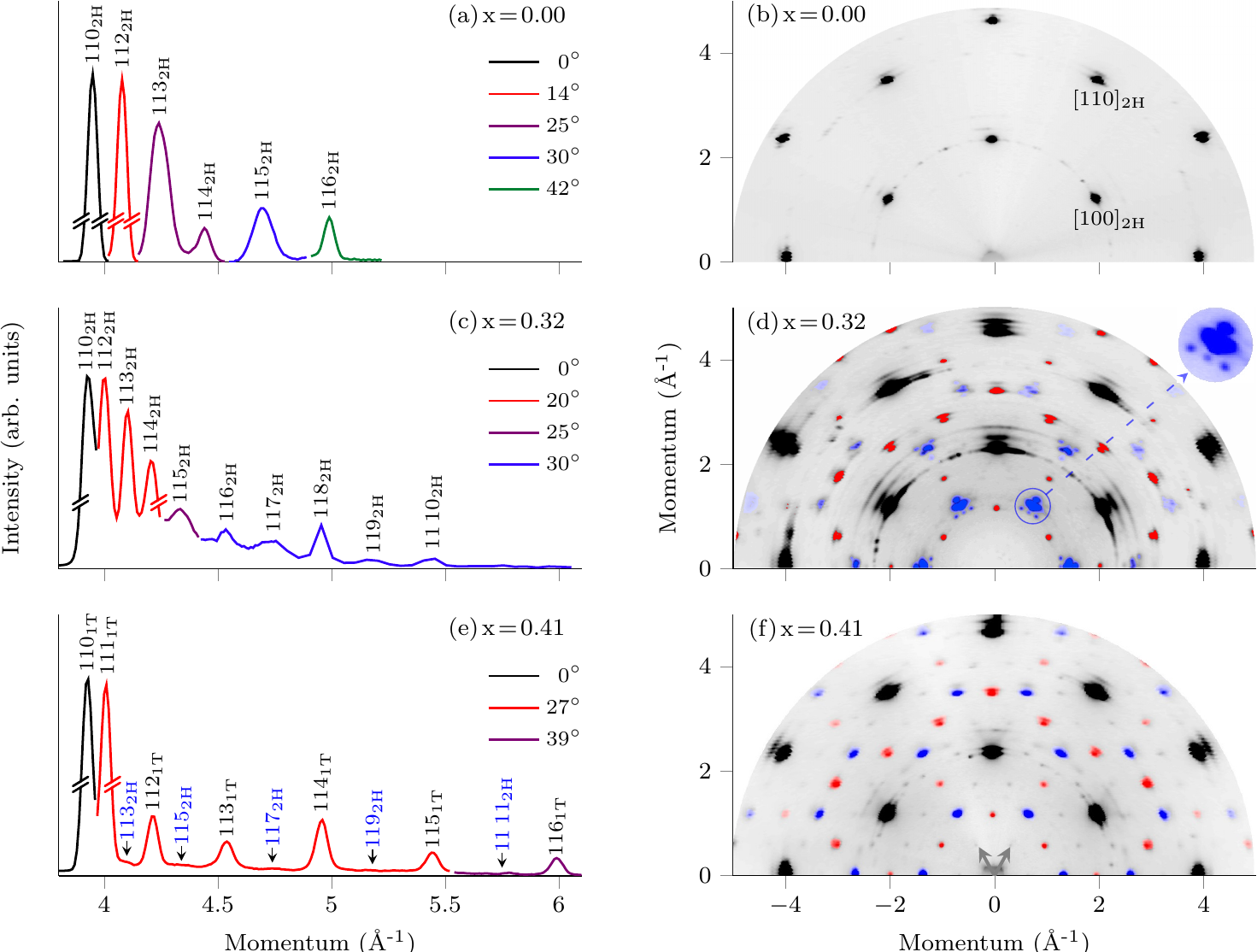}
	\caption{[(a), (c) and (e)] Diffraction peaks for the first [11$w$] reflexes of pristine and K-intercalated MoS$_2$. The angles labeling the curves refer to the polar angles used to rotate the lattice planes associated with the respective $w$ values in the spectrometer's plane of momentum transfer. The subscripts behind the lattice point indices refer to the respective polytype.
	[(b), (d) and (f)] Associated diffraction patterns in the $ab$ plane. All data were measured using elastic electron scattering at $T=\SI{300}{\K}$.}
	\label{fig:Diffraction}
\end{figure*}

To observe the changes in the crystals structure due to the potassium intercalation, we measured the diffraction patterns for various doping levels. Figures \ref{fig:Diffraction}(a), \ref{fig:Diffraction}(c) and \ref{fig:Diffraction}(e) display the diffraction peaks for a number of [11$w$] reciprocal lattice points. As described in Sec. \ref{sec:Experiment}, they were generated by aligning the lattice point of the respective plane (labeled by $w$) with the momentum transfer plane of the instrument by changing the polar angle allowing the calculation of the lattice parameter in the $c$ direction. The associated in-plane diffraction patterns are presented in Figs. \ref{fig:Diffraction}(b), \ref{fig:Diffraction}(d) and \ref{fig:Diffraction}(f).

Except for a few weak reflexes, the undoped material exhibits the hexagonally ordered diffraction peaks in the $ab$ plane [Fig. \ref{fig:Diffraction}(b)] that are typical for $2H$-MoS$_2$ [see Fig. \ref{fig:CrystalStructures}(k) for the matching simulated diffraction pattern]. The real-space lattice parameters of $a=\SI{3.18}{\angstrom}$ and $c=\SI{12.43}{\angstrom}$, derived from the data in Figs. \ref{fig:Diffraction}(a) and \ref{fig:Diffraction}(b) (summarized in Table I in the Appendix), agree well with values published by others. 
\cite{Selwyn_SolidStateIonics_1987_22_4_337, Golubnichaya_Izv.Akad.NaukSSSRNeorg.Mater._1979_15__1467, Murray_JournalofAppliedCrystallography_1979_12_3_312, Bromley_JournalofPhysicsC-SolidStatePhysics_1972_5_7_759, Wildervanck_ZeitschriftfuranorganischeundallgemeineChemie_1964_328_5-6_309, Swanson_StandardX-raydiffractionpowderpatterns_1955, Schoenfeld_ActaCrystallographicaSectionB_StructuralScience_1983_39_4_404, Lucovsky_PRB1973_3859, El-Mahalawy_JournalofAppliedCrystallography_1976_9_5_403, Pina_JournalofCrystalGrowth_1989_96_3_685, Agarwal_MaterialsResearchBulletin_1985_20_3_329,Casalot_Ann.Chim._1986_11__30921}

At a potassium concentration of $x=0.32$, the in-plane diffraction map [Fig. \ref{fig:Diffraction}(d)] still displays the main Bragg spots in almost the same positions as before as well as a number of new features that appear to create hexagonal structures around the main peaks. 
The reflexes colored in red together with the black main peaks form a $\frac{1}{2}a^* \times \frac{1}{2}a^*$ pattern associated with a $2a \times 2a$ real space superstructure [see simulated pattern in Fig. \ref{fig:CrystalStructures}(n)]. Such a structure has also been detected for Na$_{0.25}$MoS$_2$.\cite{Huang_AMI2017_1700171} However, it could also arise from three twinned $2a \times a$ lattices as described above. This is an indication that a transition to a $1T'$ or $1T''$ phase has already occurred in at least some parts of the sample. The blue tinted reflexes in Fig. \ref{fig:Diffraction}(d) consist of clusters of several features. One such cluster is magnified in the figure's inset. In conjunction with the black main spots, they are arranged in a way that corresponds to a $\sqrt{3} \times \sqrt{3}$ superstructure [see the simulated pattern in Fig. \ref{fig:CrystalStructures}(o)]. However, given that they are groups of small peaks instead of single reflexes, they are probably not related to a $1T'''$ structure and of different origin than the $2a \times 2a$ pattern. It is more likely that they arise from the potassium ordering in the Van der Waals gaps. 
The main peaks themselves [colored black in Fig. \ref{fig:Diffraction}(d)] are broader compared to the ones for undoped MoS$_2$, smeared out along the paths of equal momentum transfer and show radial multiplicity towards higher momentum transfer values. The circular smearing out is presumably the result of the misalignment of the in-plane crystal structure as a result of the incomplete structural transformation process and the waviness of the crystal layers due to an uneven potassium distribution. This waviness also causes reciprocal lattice points from adjacent lattice planes to satisfy the Bragg condition concurrently 
leading to the radial multiplicity. 

The reflexes in the $c$ direction [Fig. \ref{fig:Diffraction}(c)] can still be assigned to the $2H$ phase but they show an expansion of the out-of-plane lattice constant by $31\%$ ($c=\SI{16.28}{\angstrom}$) signified by the reduction of the peak distances. Based on the data in subfigure (c), it is not possible to distinguish between $1T$ and $2H$ polytypes because the $1T$ diffraction peaks would be in the same location as every other $2H$ peak. For the same reason, no statement can be made to say if the intercalation process leads to staging of the potassium atoms as was observed for the embedding of sodium\cite{Wang_AN2014_11394}. However, in conjunction with the in-plane diffraction map, it can be concluded that the sample consists of a mix of $2H$ and $1T$ phases with uneven potassium distribution. The amount of alkali metal was insufficient to penetrate the whole crystal uniformly. 

After further increasing the potassium concentration to $x=0.41$, every $2H$-related Bragg peak with an odd Miller index in the $c$ direction virtually disappeared from the data depicted in Fig. \ref{fig:Diffraction}(e). The locations where those peaks should have been show only very weak remnants and are indicated by the blue indices in the plot. The remaining features can be indexed as $1T$-peaks. Consequently, the material has essentially transitioned to one or more of the $1T$ phases with only one crystal layer per unit cell at $x=0.41$. 
This is similar to the intercalation level of $x=0.5$ at which the transition occurs in Na\textsubscript{x}MoS$_2$\cite{Wang_AN2014_11394} and of $x=0.4$ for which it was predicted to take place in Li\textsubscript{x}MoS$_2$\cite{Enyashin_CaTC2012_13}.
The fact that the unit cell is reduced to one molecular layer also demonstrates that there is no staging of the potassium ions at this concentration in the $c$ direction. 
The peak separation correlates to an out-of-plane lattice parameter of $c=\SI{8.23}{\angstrom}$ (see Table I in the Appendix), a $\SI{32}{\%}$ lattice expansion per layer. Increases of $33.6-\SI{34.9}{\%}$ for $x=0.4$\cite{Somoano_TheJournalofChemicalPhysics_1973_58_2_697, Ren_NR2017_1313} and $\SI{31.7}{\%}$ for $x=0.6$\cite{Ruedorff_Chimia_1965_19__489} have been observed by others.

The in-plane diffraction spots appear less smeared out compared to the lower intercalation level [Fig. \ref{fig:Diffraction}(f)] demonstrating that the crystal has become more ordered. The $2a \times 2a$ (or twinned $2a \times a$) superstructure is retained (red colored pattern in conjunction with black main peaks) pointing to a $1T'$ or $1T''$ phase. Single peaks associated with a $\sqrt{3}a \times \sqrt{3}a$ superstructure (blue colored pattern in conjunction with black main peaks) have replaced the multi-peak clusters. This type of superstructure in combination with a $2a \times 2a$ structure has not been observed in potassium doped bulk single crystal Mo2$_2$ before. They could be the result of the formation of domains containing another kind of crystal distortion such as the $1T'''$ polytype which assumes this superstructure [see Fig. \ref{fig:CrystalStructures}(o)]. However, given the evolution of those features, they are more likely related to the ordering of the alkali metal ions throughout the crystal. This was suggested for Na\textsubscript{x}TiS$_2$\cite{Hibma_PB1980_136} which resulted in a $\sqrt{3} \times \sqrt{3}$ structure among others. That conclusion is also supported by the appearance of related but very faint reflexes forming a $2\sqrt{3} \times 2\sqrt{3}$ superstructure. The associated reciprocal lattice vectors are shown as gray arrows in Fig. \ref{fig:Diffraction}(f). Previous diffraction studies on K$_{0.7}$MoS$_2$ identified the same $2a \times 2a$ superstructure\cite{Wypych_JoSSC1999_430} but neither the $\sqrt{3}a \times \sqrt{3}a$ nor the $2\sqrt{3} \times 2\sqrt{3}$ structures. 
The in-plane lattice constant associated with the black main spots is $\SI{3.20}{\angstrom}$, a mere increase of $\SI{0.6}{\%}$ compared to a $\SI{1.4}{\%}$ expansion measured for $x=0.4$ by Somoano \textit{et al.}\cite{Somoano_TheJournalofChemicalPhysics_1973_58_2_697} Overall, the volume of the unit cell changed to $\SI{72.98}{\cubic\angstrom}$.

\subsection{Energy-Loss Spectra}
\begin{figure}
	\includegraphics [width=0.48\textwidth]{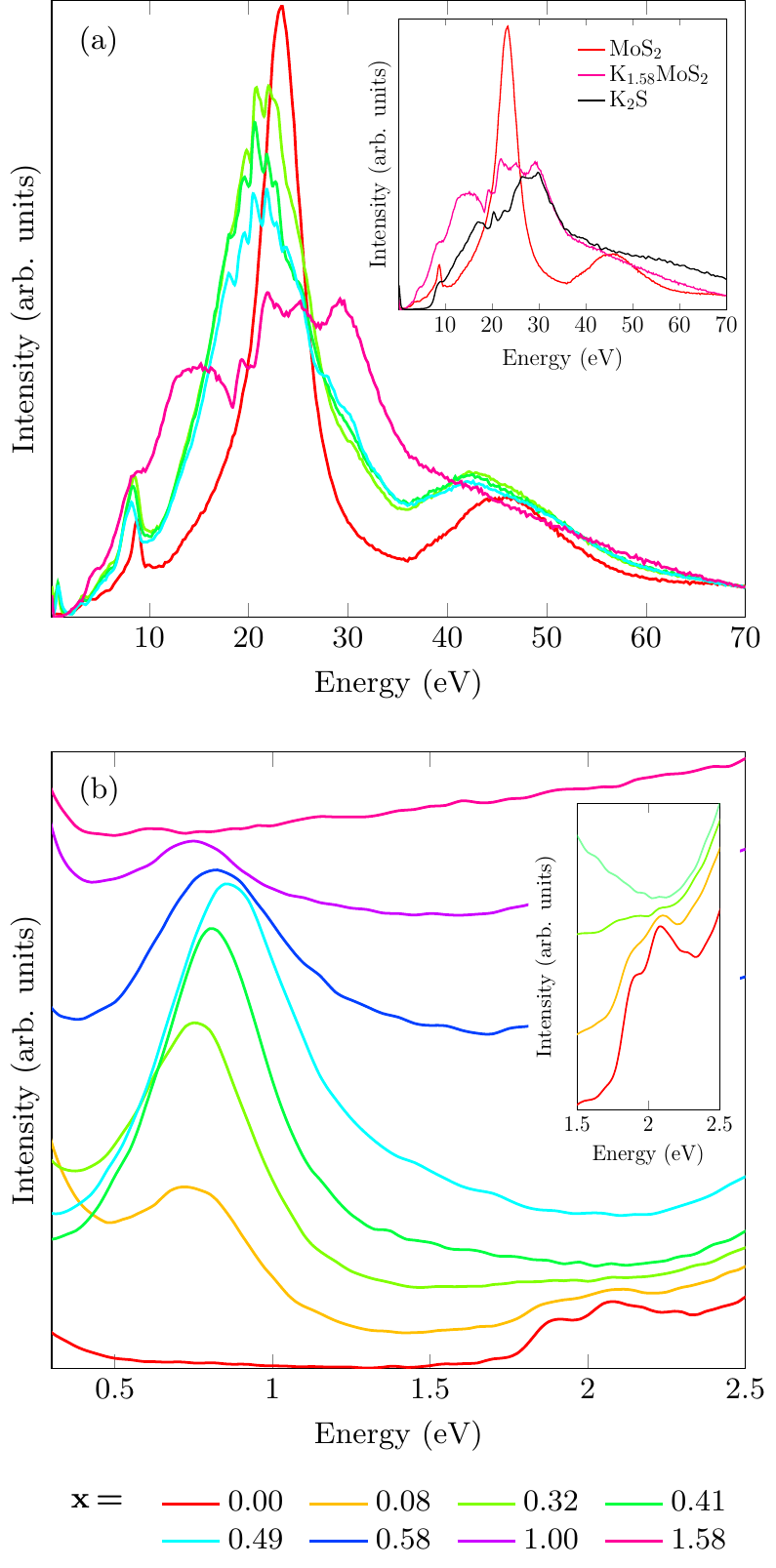}
	\caption{(a) EELS spectra for pristine MoS$_2$ and selected intercalation levels measured with a momentum transfer of $\SI{0.1}{\per\angstrom}$ parallel to the $\Gamma K$ direction at T\,=\,$\SI{300}{\K}$. The spectra were normalized at $\SI{70}{\eV}$. (b) Magnification of the spectra for the low-energy region. The inset is a rescaled plot of the energy range around the direct excitons amplifying the vanishing of the excitons at increasing intercalation levels.}
	\label{fig:DopingSteps}
\end{figure}

The energy-loss spectra measured with a momentum transfer of $|\vec{q}|=\SI{0.1}{\per\angstrom}$ in the $\Gamma$K direction for selected intercalation levels are displayed in Fig. \ref{fig:DopingSteps}. The data for the $\Gamma$M direction are not presented because they are equivalent due to the strong isotropy at low momentum transfer values. The major feature at $\SI{23.3}{\eV}$ in the spectrum of pure MoS$_2$ [red plot in Fig. \ref{fig:DopingSteps}(a)] is a reflection of the volume plasmon, the collective oscillation of all valance electrons. The smaller peak at $\SI{8.8}{\eV}$ originates from the excitation of those electrons that do not participate in the covalent bonds.\cite{Liang_PhilosophicalMagazine_1969_19_161_1031} 
The wide feature between $\SI{35}{\eV}$ and $\SI{55}{\eV}$ is a mix arising from the stimulation of the Mo $4p$ states at $\sim\!\SI{48}{\eV}$ and the effects of multiple scatting. 

With increasing intercalation level, the volume plasmon at $\SI{23.2}{eV}$ becomes broader and more jagged as a result of the excitations of the introduced K 3p electrons around $\SI{18}{\eV}$. At very high potassium concentrations, the shape of this volume plasmon changes significantly as sulfur atoms react with potassium to form potassium sulfide [magenta plot in Fig. \ref{fig:DopingSteps}(a)]. A comparison of the measured spectra of pure and heavily intercalated MoS$_2$ as wells as K$_2$S is shown in the inset for Fig. \ref{fig:DopingSteps}(a). The effects of this chemical reaction can also be seen in the K $2p$ core level spectra provided in Fig. \ref{fig:CoreLevels}. The $2p_{1/2}$ as well as the $2p_{3/2}$ peaks split at high doping levels due to the presence of unbound potassium in the Van der Waals gaps and the bound potassium in K$_2$S. Such a partial conversion to sulfide salts at high alkali metal loads has been described before for K,\cite{Somoano_TheJournalofChemicalPhysics_1973_58_2_697, Zhang_Nl2015_629} Na,\cite{Wang_AN2014_11394} and Li.\cite{Cheng_An2014_11447, Huang_SCC2018_222}

The low-energy spectrum for pristine MoS$_2$ [red plot in Fig. \ref{fig:DopingSteps}(b)] is defined by an energy gap followed by two excitations of excitonic nature\cite{Habenicht_Phys.Rev.B_2015_91__245203, Habenicht_JoPCM2018_205502} at 1.95 and $\SI{2.15}{eV}$ testifying to the material's semiconducting nature. 
As potassium is introduced, the exciton peaks begin to fade away until they are completely gone at $x=0.41$. This evolution is singled out in the inset for Fig. \ref{fig:DopingSteps}(b). It is a clear signal that the band gap disappears and that a semiconductor-to-metal transition has occurred. Such a conversion is also known for potassium intercalated WSe$_2$\cite{Ahmad_JoPCM2017_165502}, WS$_2$\cite{Ohuchi_L1989_439}, HfS$_2$ and HfSe$_2$.\cite{Habenicht_2020_} Density functional theory calculations for the latter two materials confirm this transition.\cite{Habenicht_2020_}
\begin{figure}
	\includegraphics [width=0.5\textwidth]{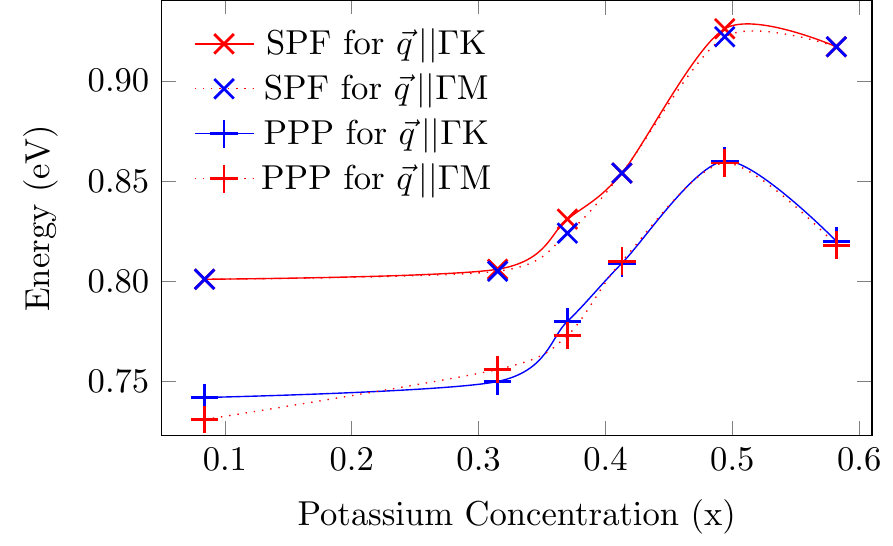}
	\caption{Screened charge carrier plasmon frequency and energy position of the plasmon peak (PPP) as a function of the average potassium intercalation level across the whole sample extracted from the energy-loss spectra measured in the $\Gamma$K and $\Gamma$M directions. The solid and dotted lines serve as a guide for the eye.}
	\label{fig:DopingSteps_Disp}
\end{figure}
Moreover, a new feature forms in the vicinity of $\SI{0.74}{\eV}$, the energy region where the band gap used to be. Its intensity increases as intercalation levels rise and decreases for $x>0.6$ until it is almost vanished at $x=1.58$ as a consequence of the decomposition to K$_2$S. The fact that this peak develops in the energy region close to $\SI{1}{\eV}$ and as a consequence of the electron transfer from potassium leading to the conversion of the material to a metal implies that the feature is a charge carrier plasmon. Such plasmons have also been observed in potassium intercalated WSe$_2$\cite{Ahmad_JoPCM2017_165502}, HfS$_2$ and HfSe$_2$\cite{Habenicht_2020_} which are semiconducting in their pristine state, too. 

The plasmon energy peak positions (PPP) were extracted from the loss spectra and plotted against the potassium concentrations in Fig. \ref{fig:DopingSteps_Disp}. The figure also displays the screened plasmon frequencies. They represent the plasmon energy peak positions after eliminating the quasielastic line from the spectra, which is an experimental artifact around $\SI{0}{\eV}$. The elimination procedure is described in Ref. \citenum{Schuster_PRB2009_45134}. The plots show that the plasmon peak positions and the screened plasmon frequencies behave very similarly except for an energy shift. The screened plasmon frequencies, in turn, follow the behavior of the unscreened plasmon frequencies  $\omega_p$ as shown in Ref. \citenum{Habenicht_2020_}. Consequently, the shifts in the spectral plasmon peak positions closely match the change in $\omega_p$. According to Eq. (\ref{equ:PlasmonFrequDrude}), the screened plasmon frequency depends on the density of the conduction electrons $n$. Therefore, the changes in the plasmon peak position can also be ascribed mainly to changes in the number of conduction electrons per unit volume. According to Fig. \ref{fig:DopingSteps_Disp}, the plasmon peak position remains relatively constant up to an average potassium concentrations of $x\approx 0.3$. This indicates that the conduction electron density up to this alkali metal load does not change significantly. Given that the conduction electrons are provided by the intercalated potassium, the K-concentration actually must stay constant. The cause for this behavior is likely that potassium stoichiometries of less than 0.3 are thermodynamically unstable. If insufficient potassium is provided in the intercalation process to achieve this stoichiometry across the whole sample, domains form that assume  the minimum required potassium concentrations while other crystal regions remain pristine. Those domains of constant potassium stoichiometry increase in volume if more potassium is added until they comprise the whole sample. This phenomenon has also been observed and computationally analyzed for potassium intercalated HfS$_2$ and HfSe$_2$.\cite{Habenicht_2020_} It should be pointed out that up to $x \approx 0.3$, the calculated potassium concentrations $x$ represent the average across the doped and undoped sample regions. For doping stoichiometries between $\sim 0.3$ and $\sim 0.5$, the plasmon peak positions increase smoothly (Fig. \ref{fig:DopingSteps_Disp}) because those concentration levels are thermodynamically stable. The energy locations of the plasmon peak decreases for $x > 0.5$ due to the decomposition of the material. Given the structural and electronic similarities, it is surprising that the K-induced plasmon in WSe$_2$ retains its energy position of $\SI{0.97}{eV}$ independent of the alkali metal concentration pointing towards the formation of a fixed stoichiometric phase.\cite{Ahmad_JoPCM2017_165502} Such a phase creation was also observed in K$_x$CuPc\cite{Flatz_TJocp2007_214702} and K$_2$MnPc\cite{Mahns_TJocp2011_194504}. On the other hand, TaSe$_2$, TaS$_2$, NbSe$_2$ and NbS$_2$, which are innately metallic TMDCs, allow for a steady potassium intercalation.\cite{Koenig_EL2012_27002, Mueller_PRB2016_35110}

The plasmon peak positions and the screened plasmon frequency also change as a function of momentum transfer ($\vec{q}$). They are a reflection of the unscreened plasmon frequency $\omega_p$ damped by single particle excitations. The general dispersion behavior of the unscreened plasmon frequency is roughly the same as that of the screened one and the PPP, except for an energy shift.\cite{Mueller_PRB2017_75150} Therefore, we can take the dispersion of the PPP as a proxy for that of the unscreened plasmon frequency.
In an ideal metal, this frequency $\omega_p(|\vec{q}|)$ increases approximately quadratically with momentum:\cite{Nolting_2018_695, Nozieres_PR1959_1254}
\begin{equation}
	\omega_p(|\vec{q}|)\approx\omega_p(0)+\frac{3}{10}\frac{\hbar^2|\vec{k}_f|^2}{m^2_e\omega_p(0)}|\vec{q}|^2+O(|\vec{q}|^4).
	\label{equ:PlasmonDisp}
\end{equation}
$\hbar$ is the Planck constant and $\vec{k}_f$ is the Fermi wave vector. 
This quadratic dispersion behavior has actually been observed in experiments.\cite{Nuecker_PRB1991_7155, Grigoryan_PRB1999_1340} However, some metallic TMDCs such as TaSe$_2$, TaS$_2$, NbSe$_2$ and NbS$_2$ exhibit a negative energy-momentum relation.\cite{Schuster_PRB2009_45134, Wezel_PRL2011_176404, Mueller_PRB2016_35110} Negative dispersion has been attributed to charge density wave instabilities or fluctuations\cite{Wezel_PRL2011_176404}, intraband transitions within the conduction bands\cite{Cudazzo_PRB2012_75121, Faraggi_PRB2012_35115, Cudazzo_NJoP2016_103050} as well as interband transitions \cite{Felde_PRB1989_10181, Kimura_JotPSoJ2015_84701}. The dispersion of the above-mentioned metallic TMDCs becomes positive and is largely linear upon potassium intercalation.\cite{Mueller_PRB2016_35110, Koenig_PRB2013_195119}

\begin{figure}
	\includegraphics [width=0.5\textwidth]{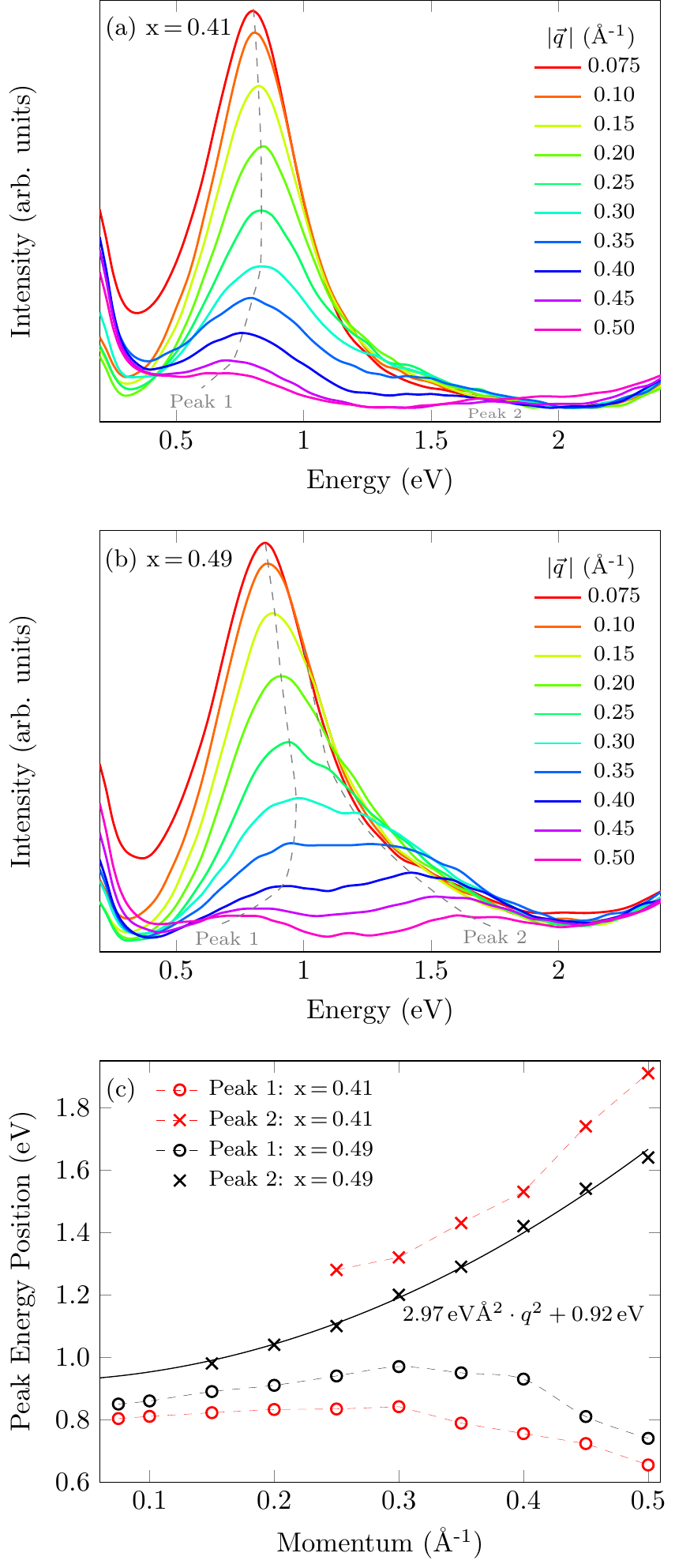}
	\caption{[(a) and (b)] EELS spectra of MoS$_2$ with potassium concentrations of $x=0.41$ and $x=0.49$, respectively, measured for the indicated momentum transfer values ($|\vec{q}|)$ in the $\Gamma$K direction at $T=\SI{300}{\K}$. The dashed lines trace the peak positions. (c) Energy-momentum dispersions for the peaks shown in (a) and (b). The solid black line outlines the quadratic fit of the dispersion. The dashed lines serve as a guide for the eye.}
	\label{fig:Dispersion}
\end{figure}

The experimental loss spectra for various $|\vec{q}|$ values measured in $\Gamma$K direction are presented in Fig. \ref{fig:Dispersion}(a) and \ref{fig:Dispersion}(b) for the intercalation levels 0.41 and 0.49, respectively. The data for the $\Gamma$M direction are virtually identical. For x\,=\,0.41, the position of the peak maximum [labeled Peak~1 in Fig. \ref{fig:Dispersion}(a)] slightly shifts to higher energies from $\SI{0.8}{eV}$ at $\SI{0.075}{\per\angstrom}$ to $\SI{0.84}{eV}$ at $\SI{0.30}{\per\angstrom}$. The process reverses for higher momentum transfer values so the peak position changes gradually down to $\SI{0.65}{eV}$ at $\SI{0.5}{\per\angstrom}$. Fig. \ref{fig:Dispersion}(c) depicts this dispersion relation by plotting the PPP versus the associated momentum transfer values. Moreover, for $|\vec{q}|=\SI{0.25}{\per\angstrom}$, a second, weaker peak [labeled Peak~2 in Fig. \ref{fig:Dispersion}(a)] appears at $\SI{\sim1.3}{eV}$ and disperses to higher energies. 
For a potassium level of 0.49, Peak~1 is again present though shifted to somewhat higher energies. It shows a similar dispersion behavior as before with an energy maximum at $\SI{0.3}{\per\angstrom}$. On the other hand, Peak~2 is more pronounced and can be distinguished already at lower momentum transfers compared to the previous intercalation stage. This feature exhibits a positive dispersion over the whole momentum range. A regression analysis arrived at a quadratic dispersion with the parameters shown next to the plot in Fig. \ref{fig:Dispersion}(c).

The question arises if both peaks represent charge carrier plasmons or only one of them. The positive, almost linear dispersion of Peak~1 up to $\SI{0.3}{\per\angstrom}$ resembles that found in the potassium doped metallic TMDCs such as TaSe$_2$ while the quadratic energy-momentum relation of Peak~2 agrees to the expectation for a metal reflected in Eq. (\ref{equ:PlasmonDisp}).
Because there can be only one charge carrier plasmon in a homogeneous material, two of them would require two regions of the sample in which the parameters of Eq. (\ref{equ:PlasmonFrequDrude}) are distinctly different. This is unlikely the case for the investigated samples not only based on the structural analysis in Sec. \ref{sec:CrystalParameters} but also the fact that the spectral feature forms one peak at low momentum transfer values and only separates at higher $\vec{q}$. If there were more than one domain with differing charge carrier concentrations $n$, the plasmon frequencies should be in different energy positions already at low $\vec{q}$ according to Eq. (\ref{equ:PlasmonFrequDrude}). Consequently, the concurrent appearance of two plasmons can be ruled out. Given that K-intercalated WSe$_2$, which has a close resemblance to MoS$_2$, shows a positive but quasilinear dispersion,\cite{Ahmad_JoPCM2017_165502} Peak~2 most likely reflects the energy-momentum relation of the charge carrier plasmon. Peak~1 probably arises from interband excitations. This conclusion is also supported by \textit{ab initio} calculations on HfS$_2$ and HfSe$_2$ performed by some of the authors.\cite{Habenicht_2020_}
\subsection{Unscreened charge carrier plasmon frequency}\label{sec:PlasmonFrequency}
\begin{figure}
	\includegraphics [width=0.49\textwidth]{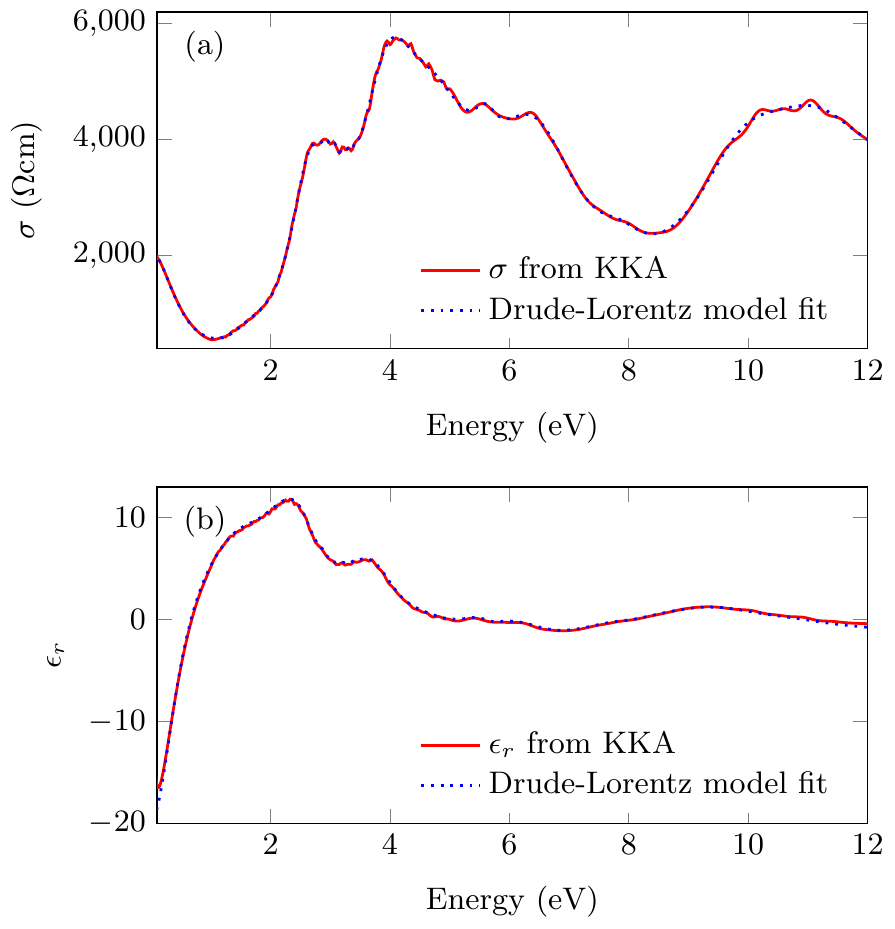}
	\caption{(a) Optical conductivity and (b) real part of the dielectric function based on the Kramers–Kronig analysis of the electron energy-loss spectrum (solid red line) and its fitting with the Drude-Lorentz model (dotted blue line).}
	\label{fig:KKA-Fits}
\end{figure}

To obtain the unscreened charge carrier plasmon frequency, it is necessary to separate the plasmon from the single particle excitations in the spectrum. For that purpose, we fitted the Drude-Lorentz model to the optical conductivity $\sigma(\omega)$ [$=\epsilon_0\omega\epsilon_{im}(\omega)$] and the real part of the dielectric function $\epsilon_r(\omega)$ derived from the EELS spectra via Kramers-Kronig analysis (KKA). 
The loss spectrum of K\textsubscript{0.41}MoS$_2$ measured in the $\Gamma$K direction in the energy range up to $\SI{70}{eV}$ was used as the basis of the analysis. The elastic line was removed from the data as well as the effects of multiple scatting according to the methods laid out in Ref. \citenum{Fink_AEEP_1989_75__121} and \citenum{Schuster_PRB2009_45134}. The normalization involved in the KKA was based on the assumption that the material is metallic. The resulting optical conductivity and the real part of the dielectric function are plotted in Fig. \ref{fig:KKA-Fits}. 
The parameters in Eq. (\ref{equ:E_Imag}) were obtained by fitting a Drude oscillator and 14 Lorentz oscillators to the optical conductivity curve up to $\SI{12}{eV}$. The use of this number of oscillators was sufficient as the fitted $\omega_{p}$ did not change any further for larger quantities. The resulting unscreened plasmon frequency was $\SI{2.78}{eV}$. The whole set of fitted parameters is listed in Table \ref{tab:FitParameters} in the Appendix. The function generated from those values is plotted as a dotted line in Fig. \ref{fig:KKA-Fits}(a) indicating that the Drude-Lorentz model provides a good representation of $\sigma(\omega)$ derived from the KKA. It highlights that the optical conductivity up to $\sim\!\SI{1}{eV}$ is largely due to the charge carrier plasmon (Drude oscillator). For energetically higher regions $\sigma(\omega)$ is a consequence of excitations of bound single particles (Lorentz oscillators). $\epsilon_{\infty}$ was determined based on Eq. (\ref{equ:E_Real}) by applying the parameters from Table \ref{tab:FitParameters} to the real part of the dielectric function from the KKA and allowing $\epsilon_{\infty}$ to vary. We found $\epsilon_{\infty}=1.56$.

\section{Calculation of intercalation level}\label{sec:IntercalationLevel}
Unfortunately, we had no means to determine the intercalation level directly. Instead, it may be estimated by multiplying the number of conduction electrons per unit volume $n$ by the volume of the unit cell provided that the sample (1) is homogeneous, (2) the conduction electrons are donated by the intercalated potassium atoms and (3) that each atom provides one electron.
The conduction electron density can be computed by rearranging Eq. (\ref{equ:PlasmonFrequDrude}) for $n$. The unscreened plasmon frequency $\omega_{p}$ in this equation is known from the Drude-Lorentz fit described above whereas $q_0$ and $\epsilon_0$ are natural constants. The only element in Eq. (\ref{equ:PlasmonFrequDrude}) not known from the experiment or the literature is the effective electron mass $m_{e}$. For our computations, we assumed that the effective mass is approximately equal to the electron rest mass.
The validity of the calculation requires a homogeneous sample. That means it has to be completely metallic, not partially decomposed to K$_2$S and entirely transformed to a $1T$ phase. The latter prerequisite ensures that a uniform unit cell volume can be determined. 

To make sure that those requirements have been met, the calculations were performed for the intercalation level at which the excitonic peaks completely disappeared for the first time from the loss spectrum, indicating that the sample is metallic, and the diffraction patterns showed a complete phase transitions. This is the spectrum labeled $x=0.41$ in Fig. \ref{fig:DopingSteps} and the associated diffraction patterns in Fig. \ref{fig:Diffraction}(e) and \ref{fig:Diffraction}(f). Of course, the value of 0.41 was not yet known at the time of calculation. The unscreened plasmon frequency for this potassium concentration is $\omega_{p}=\SI{2.78}{eV}$ (see Sec. \ref{sec:PlasmonFrequency}). Substituting this value into Eq. (\ref{equ:PlasmonFrequDrude}) results in an electron density of $5.64\cdot10^{-3}$ electrons per Å\textsuperscript{3} ($=\SI{5.64e21}{\per\cubic\cm}$). The unit cell volume associated with that sample state is $\SI{72.98}{\cubic\angstrom}$(see Sec. \ref{sec:CrystalParameters}).
Consequently, the compound possesses 0.41 conduction electrons per unit cell so the intercalation levels is $x=0.41$ (referenced to one molybdenum atom per unit cell).

However, this approach cannot be applied to potassium concentrations where multiple polytypes coexist or K$_2$S has formed. Therefore, all other potassium concentrations were calculated based on the areas under the K $2p$ core level peaks in the loss spectra.
% to the area of this peak for $x=0.41$. 
The areas were integrated after subtracting a linear background running from the spectrum at $\SI{293.4}{eV}$ to $\SI{320}{eV}$. The use of a $\omega^{-3}$-background resulted in comparable areas. The percentage changes of the area relative to the area for $x=0.41$ were applied the potassium concentration of 0.41 to obtain the intercalation levels for the data related to the other intercalation steps. 
\begin{figure}
	\includegraphics [width=0.48\textwidth]{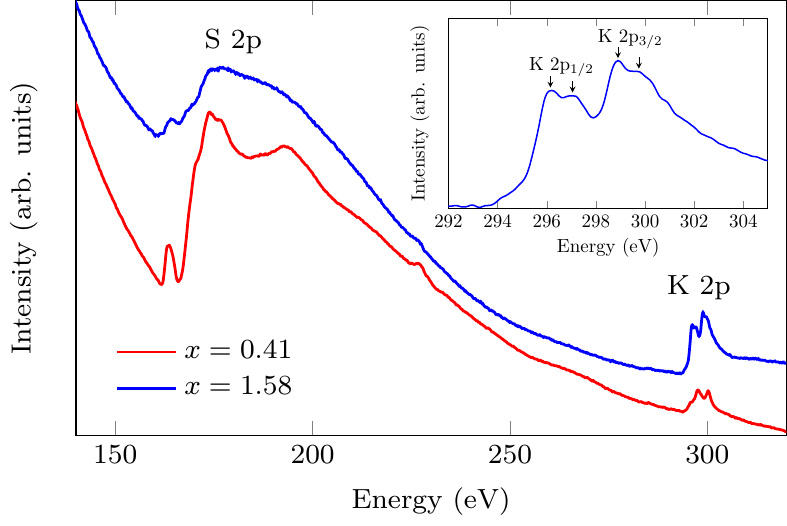}
	\caption{EELS spectra of the S $2p$ and K $2p$ core levels for the indicated intercalation levels. The inset shows a magnification of the spectra of the K $2p$ core levels for $x=1.58$. The arrows indicate the splitting of the K $2p_{1/2}$ and $2p_{3/2}$ peaks due to the formation of K$_2$S.}
	\label{fig:CoreLevels}
\end{figure}

Attempts to calculate the K concentration based on a comparison of the areas under the S $2p$ and K $2p$ core level spectra lead to unrealistic results due to the broadness of the sulfur feature (see Fig. \ref{fig:CoreLevels} for selected core level spectra).

\section{SUMMARY}
We performed electron energy-loss spectroscopy to investigate the ramifications of potassium intercalation on $2H$-MoS$_2$. Electron diffraction measurements found that at low potassium levels, $2H$ and $1T$ phases coexist in the same, unevenly doped sample. The crystals become largely homogeneous for K concentrations greater than 0.4 leading to a $2a \times 2a$ superstructure which may be the result of a twinned $2a \times a$ superstructure. Consequently, the material assumes either the $1T'$ or the $1T''$ polytype with unit cell parameters of $a=\SI{3.20}{\angstrom}$ and $c=\SI{8.23}{\angstrom}$. Moreover, a $\sqrt{3}a \times \sqrt{3}$ and a faint $2\sqrt{3}a \times 2\sqrt{3}$ superstructure are visible in the diffraction pattern which probably arise from the arrangement of the intercalated ions.
Electron energy-loss spectra showed that the exciton peaks become weaker with increasing potassium load until they completely disappear while a charge carrier plasmon developed initially at $\SI{0.74}{eV}$. The energy position of the feature remained fairly constant for intercalation levels up to $x\approx 0.3$ but shifted to higher energies for $\sim\!0.3<x<\,\sim\!0.5$. This behavior indicates that potassium stoichiometries of less than 0.3 are thermodynamically unstable leading to the formation of doped and undoped domains if the supplied amount of the alkali metal is insufficient to form the minimum stoichiometry throughout the entire sample. Potassium concentrations of more than 0.3 are thermodynamically stable. Consequently, the plasmon energy peak position increases smoothly once the whole crystal has reached the minimum doping level. For $x > 0.5$, the plasmon moved back to lower energies and faded away due to the formation of K$_2$S. The disappearance of the excitonic features and the emergence of the charge carrier plasmon indicate a complete semiconductor-to-metal transition at $x=0.41$.
For higher momentum transfer values, the plasmon peak separated into two features. One of them displayed a quasi-linear, positive dispersion up to $\SI{0.3}{\per\angstrom}$ and a negative one for higher $\vec{q}$ values and is probably related to interband transitions. The second peak, which is most likely related to the plasmon, has a positive, quadratic dispersion. 
We also used the Drude-Lorentz model to fit the optical conductivity and the real part of the dielectric function obtained via Kramers-Kronig analysis providing an unscreened plasmon frequency of $\SI{2.78}{\eV}$.

\begin{acknowledgments}
We thank E. Fischer for participating in the sample preparation and some of the EELS measurements as well as R. H\"ubel, S. Leger, M. Naumann and F. Thunig for their technical assistance. R. Schuster and C. Habenicht are grateful for funding from the IFW excellence program. A. Lubk has received funding from the European Research Council (ERC) under the Horizon 2020 research and innovation programme of the European Union (Grant Agreement No. 715620).

\end{acknowledgments}

\clearpage
\section{APPENDIX} \label{sec:AppendixA}
\begin{table}[h]
\centering
\renewcommand{\tabcolsep}{0.2cm}
\caption{Measured momentum transfer values for diffraction peak positions with Miller indices [11$w$] and the unit cell parameters in the $c$ direction calculated from those values.}
\label{tab:EnergyLevels}
\begin{threeparttable}
\begin{tabular}{cccccc}
\toprule
\toprule
\multicolumn{1}{c}{}	& \multicolumn{2}{c}{Undoped} &\multicolumn{1}{c}{} &\multicolumn{2}{c}{Doped (x=0.41)} \\ \cline{2-3} \cline{5-6}

\multicolumn{1}{c}{}	& \multicolumn{1}{c}{Momen-}	& \multicolumn{1}{c}{Calc.} &\multicolumn{1}{c}{}& \multicolumn{1}{c}{Momen-}	& \multicolumn{1}{c}{Calc.} \\

\multicolumn{1}{c}{Miller}	&	\multicolumn{1}{c}{tum}	& \multicolumn{1}{c}{Unit} &\multicolumn{1}{c}{} &	\multicolumn{1}{c}{tum}	& \multicolumn{1}{c}{Unit} \\

\multicolumn{1}{c}{Index}	&	\multicolumn{1}{c}{Transf.}	& \multicolumn{1}{c}{Cell} &\multicolumn{1}{c}{} &	\multicolumn{1}{c}{Transf.}	& \multicolumn{1}{c}{Cell}\\

\multicolumn{1}{c}{$w$\hspace{0.5pt}-value}	&	\multicolumn{1}{c}{Value}	& \multicolumn{1}{c}{Param.} &\multicolumn{1}{c}{} &	\multicolumn{1}{c}{Value}	& \multicolumn{1}{c}{Param.}\\

\multicolumn{1}{c}{ in [11$w$]} &	 \multicolumn{1}{c}{$|\vec{q}|\:(\SI{}{\per\angstrom})$}	& \multicolumn{1}{c}{c\tnote{b} \;$(\SI{}{\angstrom})$} &\multicolumn{1}{c}{} &	 \multicolumn{1}{c}{$|\vec{q}|\:(\SI{}{\per\angstrom})$}	& \multicolumn{1}{c}{c\tnote{b}\; $(\SI{}{\angstrom})$}\\
\midrule
0	&3.950	&- 		&	&3.928	&-\\
1	&\tnote{a}		&-		&	&4.003		&8.14\\
2	&4.079	&12.36	&	&4.215		&8.22\\
3	&4.221	&12.67	&	&4.545		&8.25\\
4	&4.438	&12.42	&	&4.969		&8.26\\
5	&4.695	&12.38	&	&5.470		&8.25\\
	6&4.997	&12.32	&	&6.020		&8.27\\
\midrule
\multicolumn{2}{l}{Average}	&12.43	&				&&8.23	\\
\bottomrule
\bottomrule
\end{tabular}
	\begin{tablenotes}
		\item [a] Peak could not be resolved by the measurements.
		\item [b] Calculation of unit cell parameter:\\ $c=2 \pi w/\sqrt{q_{[110]}^2-q_{[11w]}^2}$ where $q_{[11w]}$ is the momentum transfer associated with the diffraction peak of the Miller index [11$w$].
\end{tablenotes}
\end{threeparttable}
\end{table}

\begin{table}[h]
\centering
\renewcommand{\tabcolsep}{0.265cm}
\caption{Drude-Lorentz model fit parameters for the optical conductivity and the real part of the dielectric function of K\textsubscript{0.41}MoS$_2$. The calculated background dielectric constant is $\epsilon_\infty$=1.56.
}
\label{tab:FitParameters}
\begin{tabular}{ccrrr}
\toprule
\toprule
Oscillator &\multicolumn{1}{c}{Oscillator}	&	&	& \\

index $j$	&\multicolumn{1}{c}{Type}	&	\multicolumn{1}{c}{$\omega_i$ (eV)} &	\multicolumn{1}{c}{$\gamma_i$ (eV)}	&	\multicolumn{1}{c}{$\omega_{pi}$ (eV)} \\
\midrule
-&Drude &\multicolumn{1}{c}{\quad-}&0.51&2.78\\
1&Lorentz&1.69&0.90&1.39\\
2&Lorentz&2.68&0.88&4.34\\
3&Lorentz&2.93&0.18&0.68\\
4&Lorentz&3.06&0.14&0.64\\
5&Lorentz&3.24&0.25&0.96\\
6&Lorentz&3.42&0.08&0.28\\
7&Lorentz&3.98&1.28&5.90\\
8&Lorentz&4.82&1.82&5.95\\
9&Lorentz&5.60&0.52&1.53\\
10&Lorentz&6.44&1.88&6.30\\
11&Lorentz&7.87&0.95&1.64\\
12&Lorentz&10.04&2.35&7.05\\
13&Lorentz&11.03&1.20&2.66\\
14&Lorentz&12.00&2.75&7.34\\
\bottomrule
\bottomrule
\end{tabular}
\end{table}

\clearpage
%\bibliography{D:/Seafile/Literature/Dichalcogenide}
%

\end{document}